\documentclass[12pt]{article}

\usepackage[colorlinks=false]{hyperref}
\usepackage[dvips]{graphicx}
\usepackage[usenames,dvipsnames]{color}
\usepackage{epsfig}
\usepackage{pdfpages}
\usepackage{rotating}
\usepackage{amssymb}
\usepackage{amsmath,amsfonts}
\usepackage[latin1]{inputenc}
\usepackage{verbatim}
\usepackage[tableposition=top,font=small,labelfont=bf,format=hang]{caption}
\usepackage{booktabs}
\usepackage{color}

\newtheorem{thm}{Theorem}
\newtheorem{prop}{Proposition}

\newtheorem{ex}{Example}

\newcommand{\F}{\mathbb{F}}
\newcommand{\Z}{\mathbb{Z}}

\newenvironment{proof}{\noindent\textbf{Proof.}\quad}
{\hspace{\stretch{1}}%
\rule{1ex}{1ex}\\}

\addtocounter{MaxMatrixCols}{15}

\begin{document}

\title{ On isodual double Toeplitz codes\thanks{This research is supported by National Natural Science Foundation of China (61672036),
Technology Foundation for Selected Overseas Chinese Scholar, Ministry of Personnel of China (05015133), Excellent Youth Foundation of Natural Science Foundation of Anhui Province (1808085J20) and
Key projects of support program for outstanding young talents in Colleges and Universities (gxyqZD2016008).}
}

\author{
Minjia Shi, Li Xu\thanks{Minjia Shi and Li Xu, School of Mathematical Sciences, Anhui University, Hefei, Anhui, 230601,
China, {\tt smjwcl.good@163.com, 382039930@qq.com}}, 
Patrick Sol\'e\thanks{CNRS, University of Aix Marseille, Centrale Marseille, I2M, Marseille,  France, {\tt sole@enst.fr}}
}

\date{}
\maketitle

\begin{abstract}
Double Toeplitz (shortly DT) codes are introduced here as a generalization of double circulant codes.
We show that such a code is isodual, hence formally self-dual. Self-dual DT codes are characterized as double circulant or double negacirculant. Likewise, 
even DT binary codes are characterized as double circulants.
Numerical examples obtained by exhaustive search show that the codes constructed have best-known  minimum distance, up to one unit, amongst formally self-dual codes, and sometimes improve on the known values.
Over $\F_4$ an explicit construction of DT codes, based on quadratic residues in a prime field, performs equally well.
We show that DT codes are asymptotically good over $\F_q$. Specifically, we construct DT codes arbitrarily close to the asymptotic varshamov-Gilbert bound for codes of rate one half.

\end{abstract}

{\bf Keywords:} isodual codes, formally self-dual codes, double Toeplitz codes, double circulant codes\\

{\bf AMS Classification (MSC 2010):} Primary 94B05, Secondary 11C08
\section{Introduction}

Self-dual codes have received an extensive attention since the 1960's due to their many connections with invariant theory \cite{NRS}, modular forms \cite{CMF}, and combinatorial designs \cite{AM}.
A natural generalization is the class of {\it isodual} codes that is to say codes that are equivalent to their duals  \cite{AAHS,B,KL,MS}. They are, in particular, {\it formally self-dual codes},
in the sense that their weight enumerators are invariant under the MacWilliams transform \cite{BH,DGH1,DGH2,FPGH}. A powerful approach to construct isodual codes is the use of {\em double circulant codes}, whose generator matrices
consists of one identity block and a circulant block \cite{BGH}. In the present paper, we replace circulant matrices by Toeplitz matrices in that definition, and introduce double Toeplitz codes.

A matrix is called {\it Toeplitz} if every diagonal from top-left to bottom-right has the same elements. A {\it double Toeplitz} (DT) code is then a linear code with generator matrix of the form $(I~~A)$ where $I$ is an identity matrix, and $A$ a Toeplitz matrix.
Let $\textbf{a}= (t,a_1,a_2,...,a_{n-1})$ be the first row of $A$, thus the DT code reduces to a double circulant code when the first column of $A$ is $\textbf{b} = (t,a_{n-1},..., a_2,a_1)$.
Since a Toeplitz matrix satisfies $A^t=QAQ$ with $Q$ an involutive permutation matrix, it can be shown that the DT code is equivalent to its dual, and is, in particular, formally self-dual (FSD).
FSD codes have been studied extensively over $\F_2$ \cite{BH,FGPH,KP},  $\F_3$ \cite{DGH0}, $\F_4$ \cite{HaK}, and even $\F_5,$ or $\F_7$ \cite{DGH}. 
For $q=3,4,5,7$ numerical examples in short to medium lengths show the DT codes have parameters equal or up to one unit of the best-known FSD codes as per these references,
and sometimes among linear codes \cite{CT}. For $q=2$ we find four improvements on the best-known values of the minimum distance of FSD codes. Over $\F_4$ an explicit construction of DT codes, based on quadratic residues in a prime field, performs equally well.

We characterize self-dual DT codes as double circulant or double negacirculant codes. Likewise, we show
even DT binary codes are double circulants.
Further, by random coding, we can show that long DT codes are arbitrarily close in terms of  relative distance to the
Gilbert-Varshamov for linear codes of rate one half. They thus constitute a class of asymptotically good codes.

The material is arranged as follows.
Section 2 collects the notations and notions needed to follow the rest of the paper.
Section 3 studies some properties of DT codes.
Section 4 derives asymptotic bounds by random coding.
Section 5 gives some constructions of Toeplitz matrices over finite fields.
Section 6 displays some numerical examples of parameters of DT codes.
Section 7 concludes the article and points out some significant and challenging open problems.
\section{Preliminaries}
\subsection{Linear codes}

Throughout, we denote by $\F_q$ the finite field of order $q,$ where $q$ is a prime power.
The {\bf (Hamming) weight} of $x \in \F_q^N$, denoted by $W(x)$, is the number of nonzero components of $x$.
A linear $[N,k,d]$ code $C$ over $\F_q$ is a $k$-dimension subspace of $\F^N_q$ with minimal distance $d$, where the {\bf minimum distance}
is the minimum nonzero weight of the code.
The elements of $C$ are called {\bf codewords}.

A matrix $A$ of size $n \times n$ is {\bf Toeplitz matrix} if it has the following form:
$$\left(
                                            \begin{matrix}
                                            t&a_1&a_2&\cdots&\cdots&a_{n-1} \\
                                            b_1&t&a_1&\ddots&      &\vdots\\
                                            b_2&b_1&\ddots&\ddots&\ddots&\vdots\\
                                             \vdots&\ddots&\ddots&\ddots&a_1&a_2 \\
                                             \vdots&&\ddots&b_1&t&a_1\\
                                             b_{n-1}&\cdots&\cdots&b_2&b_1&t\\
                                             \end{matrix}
                                              \right),$$
where $t,a_i,b_i\in \F_q$ for $1\leq i\leq n-1.$ The vector $\textbf{a}= (t,a_1,a_2,...,a_{n-1})$ is called an upper generator vector, the vector $\textbf{b} = (t,b_1,b_2,...,b_{n-1})$ is called a lower generator vector and the ordered pair $\textbf{(a,b)}$ is called a generator vector for the matrix $A$.
 A linear code $C$ of length $2n$ is said to be {\bf double Toeplitz} (DT) if  its generator matrix $G$ is of the form
 $G=(I~~A),$ where $I$ is the identity of size $n \times n$ and $A$ is Toeplitz of the same size.
Let's take $\textbf{a}= (t,a_1,a_2,...,a_{n-1})$ for the upper generator vector, if we take $\textbf{b} = (t,a_{n-1},a_{n-2},...,a_1)$ for the lower generator vector, then $A$ is a circulant matrix, and $C$ is a (pure) double circulant code \cite{AOS}, if we take $\textbf{b} = (t,-a_{n-1},-a_{n-2},...,-a_1)$ for the lower generator vector, then $A$ is a negacirculant matrix, and $C$ is a double negacirculant code.

Recall that a matrix is called {\bf monomial} if it contains exactly one nonzero element per row and per column. Thus, a permutation matrix is monomial.
Two codes $C$ and $D$ are {\bf equivalent} if there is a monomial matrix $M$ such that $MC=D.$  
The {\bf dual} $C^ \bot$ of a code $C$ is defined w.r.t. the standard inner product.
A code is {\bf self-dual } if it is equal to its dual,
and {\bf isodual } if it is equivalent to its dual.
The {\bf weight distribution} of a code $C$ is the sequence of integers $A_i$'s for $i=0,1,...,n$, where $A_i$ is the number of codewords of weight $i.$
A code is {\bf formally self-dual} (FSD) if it has the same weight distribution as its dual. In particular, isodual codes are FSD.
A binary code is said to be  {\bf even} if the weights of all its codewords are even, and {\bf odd} otherwise.

If $C(N)$ is a sequence of codes of parameters $[N, k_N, d_N]$, the {\bf rate} $r$ and {\bf relative distance} $\delta$ are defined as
$$r=\limsup\limits_{N \rightarrow \infty}\frac{k_N}{N} \ {\rm and} \
\delta=\liminf\limits_{N \rightarrow \infty}\frac{d_N}{N}.$$
A family of codes is said to be {\bf asymptotically good} iff it contains a sequence with rate $r$ and relative distance $\delta$ such that $r\delta \neq 0.$
Recall the classical {\bf q-ary entropy function} $H_q(x),$ \cite{FECC}, of the real variable $x$ defined for $0\leq x\leq 1,$ by the formula
$$H_q(x)=x\log_q(q-1)-x\log_q(x)-(1-x)\log_q(1-x).$$

We recall the classical asymptotic Gilbert-Varshamov bound \cite{FECC}
$$r \ge 1-H_q(\delta).$$
We will use this formula for reference and comparison, and not in the proofs.




\section{Isoduality}
The following proposition is our main motivation to introduce double Toeplitz codes. Throughout the paper the exponent $^T$ denotes transposition.

{\prop (\cite[Prop 1]{CPC}) Let $A$ be a matrix satisfying $A^t=QAQ,$ with $Q$ a monomial matrix of order $2.$ The code  $C=\langle(I~~A)\rangle$ where $I$ is identity of order $n$ is an isodual code of length $2n.$}

The following theorem shows that DT codes are isodual codes.

{\thm \label{fonda} For any Toeplitz matrix $A$, there is a monomial matrix $Q$ of order $2$ such that $AQ=QA^T,$ where $Q$ is the matrix formed by reversing rows of the identity matrix. In particular, $Q$ is formed by reversing every row of the identity matrix.
Namely, the code $C=\langle(I~~A)\rangle$ is isodual.
}

\begin{proof}
 We denote by $A_i$, $B_i$ the $i-$th row and the $i-$th column of matrix $A_{n\times n}$. Because Toeplitz matrix is symmetric across its lower-left to upper-right diagonal, we have $B_i=QA_{n-i+1}^T$, for $i=1,2,...n$, then $AQ=(B_n,B_{n-1},...,B_1)=(QA_1^T,QA_2^T,...,QA_n^T)=QA^t.$
\end{proof}

The case of self-duality is characterized as follows.
{\thm \label{char}  Let $C=\langle(I~~A)\rangle$ be a DT code over $\F_q$, where

$$A=\left(
                                            \begin{matrix}
                                            t&a_1&a_2&\cdots&\cdots&a_{n-1} \\
                                            b_1&t&a_1&\ddots&      &a_{n-2}\\
                                            b_2&b_1&\ddots&\ddots&\ddots&\vdots\\
                                             \vdots&\ddots&\ddots&\ddots&a_1&a_2 \\
                                             b_{n-2}&&\ddots&b_1&t&a_1\\
                                             b_{n-1}&\cdots&\cdots&b_2&b_1&t\\
                                             \end{matrix}
                                              \right).$$
If $C$ is self-dual,
then $A$ is a circulant matrix or a negacirculant matrix, that is, $C$ is a double circulant code or a double negacirculant code.
}

\begin{proof}
 The equation $AA^T=-I$ implies that all the entries of $AA^T$ on the main diagonal are equal.
 For instance the equality of the first two diagonal entries reads off as
 $$ t^2+\sum_{i=1}^{n-1}a_i^2=b_1^2+t^2+\sum_{i=1}^{n-2}a_i^2,$$ yielding
 $b_1^2=a_{n-1}^2,$ similarly, we can get $b_i^2=a_{n-i}^2,$ i.e. $b_i=\pm a_{n-i},~i=1,2,...,n-1.$ Now we just have to prove that the signs in $b_i=\pm a_{n-i}$ are all plus or all minus.

 We denote by $A_i$ the $i-$th row of matrix $A$. Firstly, suppose these signs are neither all plus nor all minus, then there are two rows of matrix $A$, without loss of generality let's take the third and fourth rows, so that $b_2=-a_{n-2}\neq0$ and $b_3=a_{n-3}\neq0$ or $b_2=a_{n-2}\neq0$ and $b_3=-a_{n-3}\neq0$. Because $AA^T=-I$, we can get $$A_2\cdot A_3=A_3\cdot A_4=0,$$ then $$A_2\cdot A_3-A_3\cdot A_4=a_{n-3}a_{n-2}-b_2b_3=0.$$
 If $b_2=-a_{n-2}$ and $b_3=a_{n-3}$ or $b_2=a_{n-2}$ and $b_3=-a_{n-3}$, we all get $-2a_{n-3}a_{n-2}=0,$ i.e. $a_{n-3}=0$ or $a_{n-2}=0$, contradict the hypothesis.
\end{proof}


We find a property of even DT codes.

{\thm\label{dcc} An even double Toeplitz code $C$ is a (pure) double circulant code over $\F_2$.}

\begin{proof}
Let $C=\langle(I~~A)\rangle$ and the generator vector of matrix $A$ is $\textbf{(a,b)}=(t,a_1,a_2,...,a_{n-1},t,b_1,b_2,...,b_{n-1})$. Let's take any two adjacent rows of $A$, without loss of generality, take the first row $A_1=(t,a_1,a_2,...,a_{n-1})$ and the second row $A_2=(b_1,t,a_1,a_2,...,a_{n-2})$.
Because $C$ is even, each row of $A$ should have an odd weight. If $W(t,a_1,a_2,...,a_{n-2})$ is even, then $a_{n-1}=1,~b_1=1$, if $W(t,a_1,a_2,...,a_{n-2})$ is odd, then $a_{n-1}=0,~b_1=0$. So, we have $b_1=a_{n-1}$. In the same way, we can get $b_2=a_{n-2},~b_3=a_{n-3},...,b_{n-1}=a_1$, i.e. $A$ is a circulant matrix, $C$ is a double circulant code.
\end{proof}

\section{Asymptotics}
The number of DT codes of length $2n$ is important to count.
{\prop\label{omega} The number of DT codes of length $2n$ over $\F_q$ is $q^{2n-1}.$}

The easy proof is omitted. We prepare for the proof of the next theorem by a lemma from linear algebra.

{\thm \label{lambda} Let $\mathbf{u},\mathbf{v} \in \F_q^n$ and $\mathbf{u}\neq\textbf{0},$ then if $\mathbf{v}\neq\textbf{0},$ the number of DTs of length $2n$ over $\F_q$ that contains the vector $(\mathbf{u},\mathbf{v})$ is at most $q^n$, if $\mathbf{v}=\textbf{0},$ the number of DTs of length $2n$ over $\F_q$ that contains the vector $(\mathbf{u},\mathbf{v})$ is at most $q^{n-1}$.}

\begin{proof} Let $C=\langle(I~~A)\rangle$ be a DT code, the check matrix is $(-A^t~~ I)$, if $(\mathbf{u},\mathbf{v})\in C$, then $A^t\mathbf{u}^t=\mathbf{v}^t,$ where $A^t=QAQ,$ and $Q$ is as in Theorem \ref{fonda}. Letting  $(\mathbf{u},\mathbf{v})=(u_1,...,u_n,v_1,...,v_n)$, $\mathbf{u'}=Q\mathbf{u}^t=(u_n,...,u_1)^t$ and $\mathbf{v}'=Q\mathbf{v}^t=(v_n,...,v_1)^t,$ we obtain the system of $n$ equations $D\textbf{c}=\mathbf{v}'$, where $\textbf{c}=(b_{n-1},b_{n-2},...,b_1,w,a_1,a_2,...,a_{n-1})^t$,
$$D=\left(
                                            \begin{matrix}
                                            0     &\cdots &\cdots &\cdots&0      &u_n     &u_{n-1}&u_{n-2}&\cdots &u_1 \\
                                            0     &\cdots &\cdots &0     &u_n    &u_{n-1} &u_{n-2}&\cdots &u_1    &0\\
                                            0     &\cdots &0      &u_n   &u_{n-1}&u_{n-2} &\cdots &u_1    &0      &0\\
                                            \vdots&       &       &      &       &        &       &       &       &\\
                                             0    &u_n    &u_{n-1}&\cdots&\cdots &u_2     &u_1    &0       &\cdots&0\\
                                             u_n  &u_{n-1}&\cdots &\cdots&u_2    &u_1     &0      &\cdots  &\cdots&0\\
                                             \end{matrix}
                                              \right)_{n\times(2n-1)},$$
$\mathbf{u}\neq\textbf{0},$ so the rank of $D$ is $n$. If $\mathbf{v}\neq\textbf{0},$ the number of solutions of above system is $q^n$, if $\mathbf{v}=\textbf{0},$ the number of solutions of above system is $q^{n-1}.$
\end{proof}


{\thm  If  $0<\delta < H_q^{-1}(\frac{1}{2}),$ here are sequences of isodual DT codes of  relative distance $\delta.$ }

\begin{proof}
For a given $n,$  there are by Proposition 2 exactly {$\Omega_n= q^{2n-1}$} DT codes of length $2n$.
The total number of vectors over $\F_{p}^n,$ with length $n$ and Hamming weight $<d_n=\lfloor 2n\delta \rfloor,$ call it $V_n,$ is at most
\begin{equation}\label{ms}
  V_n\le{q^{2nH_q(\delta)}}
\end{equation}
By Theorem \ref{lambda}, a nonzero vector $(\mathbf{u},\mathbf{v})$ with weight $<d_n$ can be contained in at most $p^n$ such code.
If \begin{equation}\label{funda}
    \Omega_n>q^nV_n,
   \end{equation}
then there is at least one such DP code of length $2n$ with minimum distance $\ge d_n.$
 By (\ref{ms}) we see that inequality (\ref{funda})
will hold for $n$ large enough if $$ {q^{n+2nH_q(\delta)}}=o(q^{2n-1}),$$ which will hold in particular if
 $H_q(\delta)<\frac{1}{2}.$
\end{proof}

{\bf Remark:} Thus this theorem means that for every $\epsilon>0,$ there are sequences of isodual
DT codes with a relative distance $> H_q^{-1}(\frac{1}{2})-\epsilon$. Unfortunately, the method employed does not allow us to make $\epsilon=0.$
\section{Constructions}
\subsection{Self-dual codes}
Let
$$E_1=\left(
                                            \begin{matrix}
                                            0&1&0&\cdots&0 \\
                                            0&0&1&\cdots&0 \\
                                             \vdots&\vdots&\ddots&\ddots &\vdots \\
                                             0&0&0&\ddots&1\\
                                             0&0&0&\cdots&0\\
                                             \end{matrix}
                                              \right)_{n\times n},$$
and let $E_i=E_{1}^i,$ $i=2,3,...,n-1.$
In the case $q$ is a multiple of 4 plus 1, let $w$ be a square root of $-1$, then \begin{itemize}
                                                                                   \item $A=wE_i+wE_{n-i}^T$ is circulant;
                                                                                   \item $A=wE_i+(-w)E_{n-i}^t$ is negacirculant;
                                                                                   \item $A=wI$ is both circulant and negacirculant,
                                                                                  \end{itemize}

  and all of these $A$'s satisfy $AA^T=-I$.

Note that the code with generator matrix  $(I,wI)$ is Type V in the sense of the Gleason-Prange Turyn theorem \cite[Chap. 19, Th. 1]{MS}.
\subsection{Quadratic residues}
In \cite[\S 2.4]{SO} was considered the series of Toeplitz matrices defined for $n=p>2$ a prime as follows. The $a_i's$(resp $b_i'$s) are the indicator
functions of the residues  (resp. non-residues) $\pmod{p}.$ Further $t=w$ for them but we allow a general $t\in \F_4.$ We compare the DT codes obtained to the codes in \cite{CT}.

{\ex  $n=p=2$, let $\textbf{a}_2=(w,1)$, $\textbf{b}_2=(w,1)$ be the upper and lower generator vectors of Toeplitz matrix $A$, then $$G=(I~A)=\left(
                                            \begin{matrix}
                                            1&0&w&1 \\
                                            0&1&1&w\\
                                             \end{matrix}
                                              \right)$$
generates an optimal $[4,2,3]$ DT code. \\
}
{\ex $n=p=3$, let $\textbf{a}_3=(w,1,0)$, $\textbf{b}_3=(w,1,0)$ be the upper and lower generator vectors of Toeplitz matrix $A$, then $$G=(I~A)=\left(
                                            \begin{matrix}
                                            1&0&0&w&1&0 \\
                                            0&1&0&1&w&1\\
                                            0&0&1&0&1&w\\
                                             \end{matrix}
                                              \right)$$
generates a quasi-optimal $[6,3,3]$ DT code.
}
{\ex $n=p=5$, let $\textbf{a}_5=(w,1,0,0,1)$, $\textbf{b}_5=(w,1,0,1,1)$ be the upper and lower generator vectors of Toeplitz matrix $A$, then $$G=(I~A)=\left(
                                            \begin{matrix}
                                            1&0&0&0&0&w&1&0&0&1 \\
                                            0&1&0&0&0&1&w&1&0&0\\
                                            0&0&1&0&0&0&1&w&1&0\\
                                            0&0&0&1&0&1&0&1&w&1\\
                                            0&0&0&0&1&1&1&0&1&w\\
                                             \end{matrix}
                                              \right)$$
generates a quasi-optimal $[10,5,4]$ DT code.
}

{\ex $n=p=7$, let $\textbf{a}_{11}=(w,0,0,1,0,1,1)$, $\textbf{b}_{11}=(w,1,1,0,1,1,1)$ be the upper and lower generator vectors of Toeplitz matrix $A$, then $G=(I~A)$
generates a quasi-optimal $[14,7,5]$ DT code.
}

{\ex $n=p=11$, let $\textbf{a}_{11}=(w,1,0,1,1,1,0,0,0,1,0)$, $\textbf{b}_{11}=(w,1,0,0,1,0,1,1,1,1,1)$ be the upper and lower generator vectors of Toeplitz matrix $A$, then $G=(I~A)$
generates a $[22,11,7]$ DT code, a unit away from the best-known $[22,11,8].$
}

\section{Numerics}
In Tables 1-2 and 4-5, for, respectively, $q=2,3,5,7$ we denote by
\begin{itemize}
\item $d_F(q,2n)$ the highest known minimum weight of FSD codes over $\F_q$ as per \cite{BH,DGH0,DGH},

\item $d_F^\ast(q,2n)$ the highest minimum weight of FSD DT codes constructed over $\F_q.$
\end{itemize}
We put a star exponent on the entry $d_F^\ast(q,2n)$ whenever $d_F^\ast(q,2n)=d_F(q,2n).$
We write $d_F^\ast(q,2n)$ in boldface whenever $d_F^\ast(q,2n)>d_F(q,2n).$ \ \ \\

In Table 3, we denote by $d_{fsdao}(4,2n)$ the highest minimum weight of formally self-dual additive odd codes over $\F_4$ (\cite{HaK}); and by $d_{fsd}^\ast(4,2n)$ the highest minimum weight of FSD DT codes that we can find over $\F_4$.
We put a star exponent on the entry $d_{fsd}(4,2n)^\ast$ whenever $d_{fsd}(4,2n)^\ast\geq d_{fsdao}(4,2n).$
We constructed a large number of random DT codes, and the Tables collect the best found.
All the DT codes constructed in this section are FSD codes by Theorem 1.
All computations were performed in Magma \cite{M}.
It is worth mentioning that we find a series of MDS codes that differ from the construction method on the code table \cite{CT} over $\F_4$, when length $N=6$.
\begin{table}
\centering
\begin{center}
\small {Table $1$: The Highest Minimum Weight for $\F_2$} \\
\end{center}
  \begin{tabular}{ccc}
   \toprule
    Length $2n$ &$d_F^\ast(2,2n)$&$d_F(2,2n)$\\
   \midrule
     4 &2$^\ast$&2 \\
     6 &3$^\ast$&3 \\
     8 &\textbf{4}$^\ast$&3 \\
     10&4$^\ast$&4 \\
     12&4$^\ast$&4 \\
     14&4$^\ast$&4\\
     16&5$^\ast$&5\\
     18&\textbf{6}$^\ast$&5\\
     20&6$^\ast$&6\\
     22&7$^\ast$&7\\
     24&\textbf{8}$^\ast$&7\\
     26&7$^\ast$&7\\
     28&\textbf{8}$^\ast$&7\\
     30&8$^\ast$&7 or 8\\
     32&8$^\ast$&8\\
     34&8$^\ast$&8\\
     36&8$^\ast$&8\\
     38&8$^\ast$&8 or 9\\
     40&9$^\ast$&9 or 10\\
   \bottomrule
  \end{tabular}
\end{table}
\pagebreak

\begin{table}
\centering
\begin{center}
\small {Table $2$: The Highest Minimum Weight for $\F_3$} \\
\end{center}
  \begin{tabular}{ccc}
   \toprule
    Length $2n$ &$d_F^\ast(3,2n)$&$d_F(3,2n)$\\
   \midrule
     4 &3$^\ast$&3 \\
     6 &3$^\ast$&3 \\
     8 &4$^\ast$&4 \\
     10&5$^\ast$&5 \\
     12&6$^\ast$&6 \\
     14&6$^\ast$&6\\
     16&6$^\ast$&6\\
     18&6$^\ast$&6\\
     20&7$^\ast$&7\\
     22&8$^\ast$&8\\
     24&9$^\ast$&9\\
     26&8$^\ast$ &8 or 9\\
     28&8        &9 or 10\\
     30&9$^\ast$ &9, 10 or 11\\
   \bottomrule
  \end{tabular}
\end{table}
\pagebreak
\begin{table}
\centering
\begin{center}
\small {Table $3$: The Highest Minimum Weight for $\F_4$} \\
\end{center}
  \begin{tabular}{ccc}
   \toprule
    Length $2n$ &$d_{fsdao}^\ast(4,2n)$&$d_{fsdao}(4,2n)$\\
   \midrule
     4&3$^{\ast}$&$3$ \\
     6&4$^{\ast}$&$3 $\\
     8&4$^{\ast}$&$4$ \\
     10&5$^{\ast}$&$5$ \\
     12&5         &$6$ \\
     14&6$^{\ast}$&$6$ or $7$\\
   \bottomrule
  \end{tabular}
\end{table}
\begin{table}
\centering
\begin{center}
\small {Table $4$: The Highest Minimum Weight for $\F_5$} \\
\end{center}
  \begin{tabular}{ccc}
   \toprule
    Length $2n$ &$d_F^\ast(5,2n)$&$d_F(5,2n)$\\
   \midrule
     4&3$^\ast$&3 \\
     6&4$^\ast$&4 \\
     8&4$^\ast$&4 \\
     10&5$^\ast$&5 \\
     12&6$^\ast$&6 \\
     14&6$^\ast$&6\\
     16&7$^\ast$ &7\\
     18&7$^\ast$ &7 or 8\\
     20&8$^\ast$ &8 or 9\\
     22&8$^\ast$ &8, 9 or 10\\
     24&8 &9 or 10\\
   \bottomrule
  \end{tabular}
\end{table}
\pagebreak

\begin{table}
\centering
\begin{center}
\small {Table $5$: The Highest Minimum Weight for $\F_7$} \\
\end{center}
  \begin{tabular}{ccc}
   \toprule
    Length $2n$ &$d_F^\ast(7,2n)$&$d_F(7,2n)$\\
   \midrule
     4&3$^\ast$&3 \\
     6&4$^\ast$&4 \\
     8&5$^\ast$&5 \\
     10&5$^\ast$&5 \\
     12&6$^\ast$ &6 \\
     14&7$^\ast$ &7\\
     16&7$^\ast$ &7 or 8\\
     18&8$^\ast$ &8 or 9\\
     20&8 &9 or 10\\
     22&8 &9, 10 or 11\\
     24&9 &10, 11 or 12\\
   \bottomrule
  \end{tabular}
\end{table}
\section{Conclusion and open problems}
In this work we have introduced the double Toeplitz codes. These codes can be regarded as a generalization of double circulant codes.
These codes are isodual, and in particular formally self-dual. In short lengths, their parameters are optimal or quasi-optimal amongst FSD codes,
and sometimes amongst linear codes \cite{CT} .
More importantly, we could show that they are asymptotically good over $\F_q$. In fact, their relative distance satisfies
the Gilbert-Varshamov bound for linear codes of rate one half.

\end{document}